\documentstyle[prc,aps,epsfig]{revtex}

\def\lambdas{\mbox{{$\Lambda_s$}}}
\def\lambdasr{\mbox{{$\Lambda_sR$}}}

\begin{document}
\twocolumn[\hsize\textwidth\columnwidth\hsize
           \csname @twocolumnfalse\endcsname

\draft

\title{
    Coherent gluon production in very high energy heavy ion collisions
}
\author{Alex Krasnitz$^{\rm a}$,
        Yasushi Nara$^{\rm b}$
        and
        Raju Venugopalan$^{\rm b,c}$
        }

\address{
 a\ CENTRA and Faculdade de Ci\^encias e Tecnologia, Universidade do Algarve,Campus de Gambelas, P-8000 Faro, Portugal.\\ 
 b\ RIKEN BNL Research Center, Brookhaven National Laboratory,
		Upton, N.Y. 11973, U.S.A.\\
 c\ Physics Department, Brookhaven National Laboratory, Upton,
 N.Y. 11973, U.S.A.\\
}

\date{\today}
\maketitle

\begin{abstract}
The early stages of a relativistic heavy-ion collision are examined in
the framework of an effective classical SU(3) Yang-Mills theory in the
transverse plane.  We compute the initial energy and number
distributions, per unit rapidity, at mid-rapidity, of gluons produced
in high energy heavy ion collisions.  We discuss the phenomenological
implications of our results in light of the recent RHIC data.

\end{abstract}



\vskip2pc]




The Relativistic Heavy Ion Collider (RHIC) is currently colliding
beams of gold nuclei at the highest center of mass energies, per
nucleon, $\sqrt{s_{NN}}= 200$ GeV. The goal of these experiments is to
explore strongly interacting matter, in particular the quark gluon
plasma (QGP) predicted by lattice QCD~\cite{QMproc}.

The possible formation and dynamics of the QGP depends crucially on
the initial conditions, namely, the distribution of partons in each of
the nuclei {\it before} the collision. At high energies and for large
nuclei, parton distributions saturate and form a color glass
condensate (CGC). (For other pQCD based approaches, see
Ref.~\cite{pQCD} and references therein.)  

The physics of saturated
gluons in the CGC is as follows.  As the energies of the colliding
nuclei grow (equivalently $x_{Bj} \ll 1$), partons in the nuclear wave
functions multiply until they begin to overlap in phase
space. Repulsive interactions among the partons
ensure that the occupation number saturates at a value
proportional to $1/\alpha_s$. This phenomenon~\cite{GLR}, 
is characterized by a bulk scale (the
saturation scale) $\Lambda_s$, where $\Lambda_s^2$ is proportional to
the gluon density per unit area in a nucleon or nucleus. 
A simple saturation model for nucleons with $\Lambda_s^2 = \Lambda_{s0}^2
(x_0/x_{Bj})^\delta$ with $\Lambda_{s0}=1$ GeV, $x_0=3\cdot 10^{-4}$
and $\delta= 0.29$ describes well deeply inelastic scattering data at
the Hadron Electron Ring Accelerator (HERA) for $x_{Bj}<0.01$ and all
values $Q^2$ of the transverse momentum squared from $Q^2\sim 0$ up to
$Q^2=450$ GeV$^2$~\cite{HERAfit}. For nuclei, one expects  
that $\Lambda_s^2\approx \Lambda_{QCD}^2 A^{1/3}/x^\delta$.

In a heavy ion collision, the CGC ``shatters''
producing ``on shell'' gluons. In this letter, we obtain
non-perturbative expressions relating the energy and number
distributions of produced gluons to the saturation scale $\Lambda_s$ of the 
CGC. Therefore, in principle, the saturation scale $\Lambda_s$ may be 
determined from heavy ion experiments.

The CGC can be quantified in a classical effective field theory where
$\Lambda_s^2$ is the only dimensionful scale~\cite{MV}.  When
$\Lambda_s^2\gg\Lambda_{QCD}^2$ (for high energies and large nuclei),
the coupling is weak: $\alpha_s\equiv \alpha_s(\Lambda_s^2) \ll
1$. However, the occupation number is large, $\propto 1/\alpha_s \gg
1$. Thus weak coupling, classical methods are applicable and can be
used to compute the classical parton distributions of
nuclei~\cite{MV,JKMW}. Recently, renormalization group methods have
been developed which systematically incorporate quantum corrections to
the EFT~\cite{JKMW,RG}.

The classical EFT can be applied to nuclear collisions
~\cite{GyulassyMclerran,KMW,DYEA}. 
The spectrum of gluons produced when the CGC shatters 
is described by the solution of the 
classical Yang--Mills equations in the presence of two light cone 
sources, one for each nucleus, 
with initial conditions for the gauge fields given by the gauge 
fields of the two nuclei before the collision. Analytical expressions for 
classical gluon production were obtained to lowest order in the parton 
density~\cite{GyulassyMclerran,KMW,DYEA}.
However, these are infrared divergent and need to be 
summed to all orders in the parton density. This was first done numerically 
by two of us for an SU(2) gauge theory~\cite{AR99} and non-perturbative 
expressions relating the the energy~\cite{AR00} 
and number~\cite{AR01} distributions of produced gluons to the saturation
scale were obtained. Here we extend the work of 
Refs.~\cite{AR00,AR01} to an SU(3) gauge theory~\cite{Yuri}. 
Our results can thereby be compared to available and forthcoming data from 
RHIC.

Simulating the SU(3) theory is technically more difficult
than the SU(2) theory. For a comparable set of parameters, 
the SU(3) case is about an order of magnitude more challenging numerically
than the SU(2) one. The lattice
formulation of the theory is described in detail in \cite{AR99}. The 
numerical techniques we use are well-known in lattice gauge theory, with one 
notable exception. Specifically, a new procedure had to be devised in 
order to determine the initial condition for the transverse components of the
gauge fields. To this end, one must solve Eq. 37 of \cite{AR99}:
\begin{eqnarray}
z_\mu\equiv i\rm{Tr}\lambda_\mu\left[(U^{(1)}+U^{(2)} )(I + U^\dagger)
-{\rm h.c.} \right] = 0 \, , \label{IC}
\end{eqnarray}
where $U^{(1),(2)}$ are SU(3) group elements corresponding to CGC of the two
nuclei before the collision, $U$ is the sought group element to the initial
gauge field, and $\lambda_\mu$ are the Gell-Mann matrices. In other words,
an Hermitean matrix
$M(U)\equiv i[(U^{(1)}+U^{(2)} )(I + U^\dagger)-{\rm h.c.}]$ must be 
proportional to the unit matrix.
In order to solve (\ref{IC}) numerically for $U$, we first form a non-negative
function
\begin{eqnarray}
F(U)\equiv z_\mu z_\mu={\rm Tr}(M^2)-{1\over 3}({\rm Tr}M)^2\label{fu}.
\end{eqnarray}
If $U$ satisfies (\ref{IC}), $F(U)$ attains its absolute minumum, $F(U)=0$.
Next, we minimize $F(U)$ by relaxation. The relaxation equation has the form
\begin{eqnarray}
{\dot U}=-i\lambda_\mu\partial_{\gamma_\mu}
F({\rm e}^{i\gamma_\mu\lambda_\mu}U)|_{\gamma_\mu=0}U\,
\label{relax}\end{eqnarray}
where $\gamma_\mu$ are real variables, and ${\dot U}$ is the derivative with
respect to the relaxation time $t$. The explicit expression for the 
right-hand side of (\ref{relax}) is somewhat lengthy and will be presented 
elsewhere. The relaxation equation is then integrated numerically to yield the
initial condition we seek.

In this work we will determine two observables: the energy and 
the number distribution of produced gluons. In doing so, we closely follow the 
procedure developed for the SU(2) case. 
In the continuum limit the theory contains two dimensional
parameters: $\Lambda_s$ and the nuclear radius $R$. Any observable can 
therefore be expressed as a power of $\Lambda_s$,
times a function of the dimensionless product $\Lambda_s R$ and of the coupling
constant $g$.

\begin{figure}
\epsfig{file=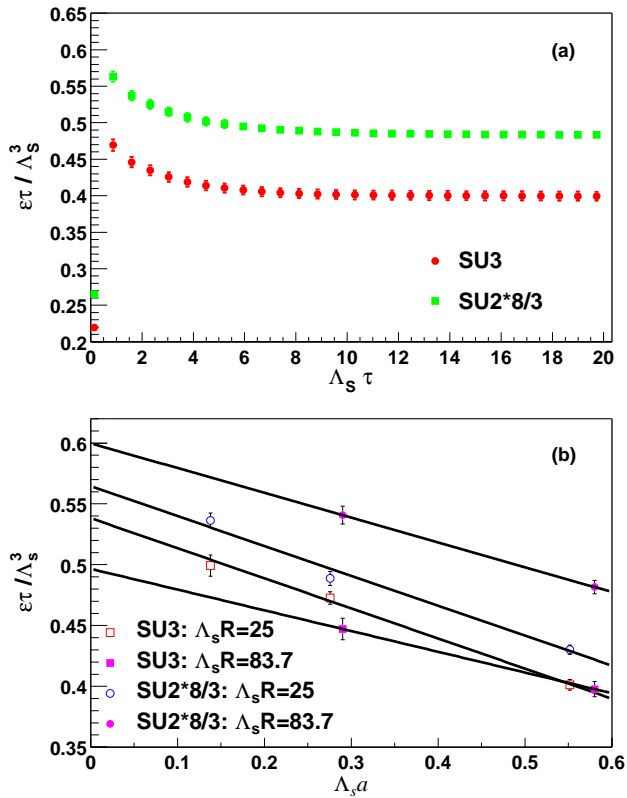,width=9cm}%
\caption{
(a) $\varepsilon\tau/\lambdas^3$ as a function of $\tau\lambdas$ for
$\lambdasr = 83.7$.
(b) $\varepsilon\tau/\lambdas^3$ as a function of $\lambdas a$ for
$\lambdasr = 83.7$ (squares) and 25(circles), where $a$ is
the lattice spacing.
Lines are fits of the form $a-bx$.
}
\label{fig:etau}
\end{figure}

For the transverse energy of gluons we get 
\begin{equation}
 {1\over \pi R^2}{dE_T\over d\eta}|_{\eta=0}
     = {1\over g^2}f_E(\Lambda_s R)\Lambda_s^3,
 \label{eq:dedy}
\end{equation}
The function $f_E$ is  
determined non-perturbatively as follows. In Figure~\ref{fig:etau} (a),
 we plot the Hamiltonian density,
for a particular fixed value~\cite{comment2} of $\Lambda_s R=83.7$ (on
a $512\times 512$ lattice) in dimensionless units as a function of the
proper time in dimensionless units. We note that in the SU(3) case, as
in SU(2), $\varepsilon\tau$ converges very rapidly to a constant
value. The form of $\varepsilon \tau$ is well parametrized by the
functional form $\varepsilon\tau= \alpha + \beta\exp(-\gamma \tau)$. Here
$dE_T/d\eta/\pi R^2 = \alpha$ has the proper interpretation of being
the energy density of produced gluons, while
$\tau_D=1/\gamma/\Lambda_s$ is the ``formation time'' of the produced
glue. 

In Figure~\ref{fig:etau} (b),
 the convergence of $\alpha$ to the continuum limit
is shown as a function of the lattice spacing in dimensionless
units for two values of $\Lambda_s R$. 
In Ref.~\cite{AR00}, this convergence to the continuum limit was 
studied extensively for very large lattices (up to $1024\times 1024$ sites) 
and shown to be linear. The trend is the same for the SU(3) results-thus, 
despite being further from the continuum limit for SU(3) (due to the 
significant increase in computer time) a linear extrapolation is justified. 
We can therefore extract the continuum value for $\alpha$. We find 
$f_E(25)=0.537$ and $f_E(83.7)=0.497$.
 The RHIC value likely lies in this 
range of $\Lambda_s R$.  The formation time 
$\tau_D=1/\gamma/\Lambda_s$ is essentially the same for SU(2)-for 
$\Lambda_s R = 83.7$, $\gamma=0.362\pm 0.023$. As discussed 
in Ref.~\cite{AR00}, it is $\sim 0.3$ fm for RHIC and $\sim 0.13$ fm for LHC 
(taking $\Lambda_s=2$ GeV and $4$ GeV respectively).

We now combine our expression in Eq.~(\ref{eq:dedy}) with our 
non-perturbative expression for the formation time to obtain a 
non-perturbative formula for the initial energy density,
\begin{equation}
\varepsilon = {0.17\over g^2}\, \Lambda_s^4
\label{eq:edens}
\end{equation}
This formula gives a rough estimate~\cite{comment4} of the initial energy density, at 
a formation time of $\tau_D = 1/\bar{\gamma}/\Lambda_s R$ where we have 
taken the average value of the slowly varying function $\gamma$ to be 
$\bar{\gamma}=0.34$.

\begin{figure}
\epsfig{file=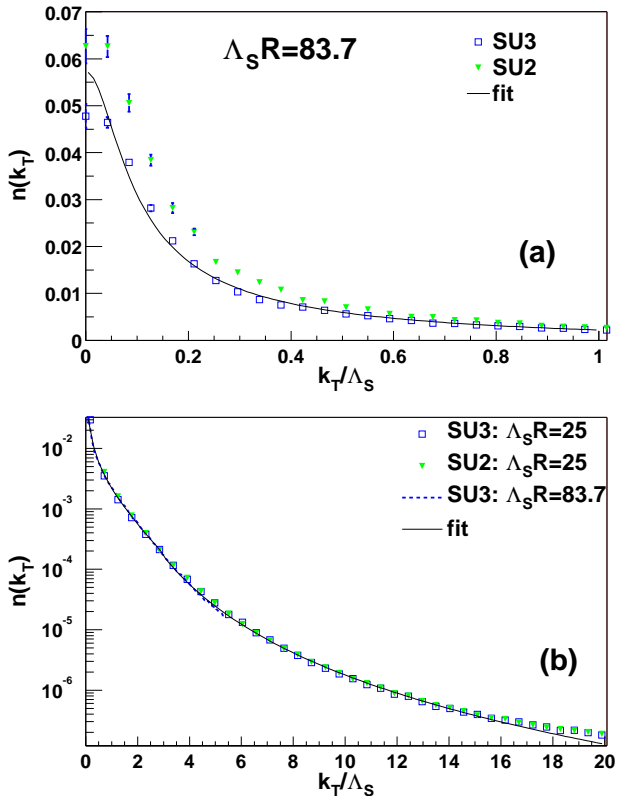,width=9cm}%
\caption{
Transverse momentum distribution of gluons, normalized to the color degrees
of freedom, $n\,(k_T) = {\tilde f}_n/(N_c^2-1)$ (see Eq.~(\ref{momdis}))
as a function of $\Lambda_SR$ for SU(3) (squares)
and SU(2) (diamonds).
Solid lines correspond to the fit in Eq.(\ref{eq:fit}).
}
\label{fig:dNdpt}
\end{figure}

To determine the gluon number per unit rapidity, 
we first compute the gluon transverse momentum distributions.
The procedure followed is identical to that described in Ref.~\cite{AR01}
-we compute the number distribution in Coulomb 
gauge~\cite{comment3}, $\nabla_\perp \cdot A_\perp =0$. 
In Fig.~\ref{fig:dNdpt}(a), we plot the normalized gluon transverse momentum
distributions versus
$k_T/\Lambda_s$ with the value $\Lambda_s R = 83.7$, together with SU(2)
result.  Clearly, we see that the normalized result for SU(3) is
suppressed relative to the SU(2) result in the low momentum region.
In Fig.~\ref{fig:dNdpt}(b), we plot the same quantity over a wider
range in $k_T/\Lambda_s$ for two values of $\Lambda_s R$. At large
transverse momentum, we see that the distributions scale exactly as
$N_c^2-1$, the number of color degrees of freedom. This is as expected 
since at large transverse momentum, the modes are nearly
those of non--interacting harmonic oscillators. At smaller momenta,
the suppression is due to non-linearities, whose effects, we have 
confirmed, are greater
for larger values of the effective coupling $\Lambda_s R$. 
The SU(3) gluon momentum distribution can be fitted by
the following function,
\begin{equation}
 {1\over \pi R^2}{dN\over d\eta d^2k_{T}}
     = {1\over g^2}{\tilde f}_n(k_{T}/\lambdas)\, ,
\label{momdis}
\end{equation}
where ${\tilde f}_n(k_{T}/\lambdas)$ is
\begin{equation}
{\tilde f}_n = \left\{
  \begin{array}{ll}
     a_1\left[\exp\left(\sqrt{k_T^2+ m^2}/ T_{\rm eff}\right) -1\right]^{-1}
                       & (k_{T}/\lambdas \leq 3) \\ \\                

    a_2\,\lambdas^4\log(4\pi k_{T}/\lambdas)k_T^{-4}
                       & (k_{T}/\lambdas > 3) \\
 \end{array} \right.
\label{eq:fit}
\end{equation}
with $a_1=0.0295$, $m=0.067\Lambda_s$, $T_{\rm eff}=0.93\Lambda_s$,
and $a_2=0.0343$.
At low momenta, the functional form is approximately that of a Bose-Einstein
distribution in two dimensions even though the underlying dynamics is that 
of classical fields. The functional form at high momentum is motivated
by the lowest order perturbative calculations~\cite{GyulassyMclerran,KMW,DYEA}.

Integrating our results over all momenta, we obtain for the gluon 
number per unit rapidity, the non-perturbative result,  
\begin{equation}
 {1\over \pi R^2}{dN\over d\eta}|_{\eta=0}
     = {1\over g^2} f_N(\Lambda_s R)\Lambda_s^2 \, .
 \label{eq:dNdy}
\end{equation}
We find that $f_N(83.7)=0.3$. 
The results for a wide range of $\Lambda_s R$ vary
 on the order of $10$\% in the case of SU(2).
%

The broad features of the CGC picture have recently been compared to 
the RHIC data~\cite{KN,JM}.
We shall here discuss the phenomenological implications of our specific 
model in light of the recent RHIC data on multiplicity and energy
distributions. The final multiplicity of hadrons~\cite{comment5} is
related to the initial gluon multiplicity by the relation
$dN^{h}/d\eta = \kappa_{inel}\,dN_i^g/d\eta$. Here $\kappa_{inel}$ is
a factor accounting for $2\rightarrow n$ gluon number changing processes
which may occur at late times beyond when the classical approach is
applicable~\cite{chem}. Moreover, if partial or full thermalization 
does occur~\cite{chem,BMSS}, the initial 
transverse energy is reduced-both 
due to inelastic collisions prior to thermalization
and subsequently due to hydrodynamic expansion-by a factor $\kappa_{work}$. 
We then have
\begin{eqnarray}
{dE_T^{h}\over d\eta}|_{\eta=0}
     &=& {\pi\over g^2}\, {1\over \kappa_{work}}\,f_E(\Lambda_s R)\,\Lambda_s 
\,(\Lambda_s R)^2\, ,\nonumber\\
 {dN^{h}\over d\eta}|_{\eta=0}
     &=& {{\pi \kappa_{inel}}\over g^2} f_N(\Lambda_s R)\,(\Lambda_s R)^2 \, .
 \label{eq:dpidy}
\end{eqnarray}

From the RHIC data at $\sqrt{s_{NN}}=130$ GeV, we have
$dN^{h}/d\eta|_{\eta=0}\sim 1000$ for central
collisions~\cite{PHOBOS00,PHENIX01,STAR01,BRAHMS01}. For $g=2$
($\alpha_s=0.33$), $\pi R^2=148$ fm$^2$, and $f_N=0.3$, we have
$\kappa_{inel} \Lambda_s^2=3.5$ GeV$^2$. Now, from Eq.~(\ref{eq:dpidy}), the
ratio $R^h=dE_T^h/d\eta/dN^h/d\eta$ is, since $f_E/f_N=1.66$,
$R^h=1.66\Lambda_s/\kappa_{work}/\kappa_{inel}$. The experimental
value~\cite{PHENIX01} for $\sqrt{s_{NN}}=130$ GeV is
$R^h=0.5$ GeV. Now, if we assume that there is no work done due to
thermalization, $\kappa_{work}=1$, we obtain from the two conditions
$\Lambda_s=1.02$ GeV and $\kappa_{inel}=3.4$ as the values that give
agreement with the data. The latter value is the maximal amount of
inelastic gluon production possible. Alternatively, if we assume
hydrodynamic work is done, one obtains
$\kappa_{work}=(\tau_f/\tau_i)^{1/3}$, where $\tau_f$ and $\tau_i$ are
the final and initial times of hydrodynamic expansion
respectively. This gives us $\kappa_{work}\approx 2$. Following the
same analysis as previously, we obtain $\Lambda_s=1.28$ GeV and
$\kappa_{inel}=2.13$. Thus, within the CGC approach, we are able to
place bounds on both the saturation scale and on the
amount of inelastic gluon production at RHIC energies. An independent 
method to extract $\Lambda_s$ directly from the data (albeit 
assuming parton-hadron duality) is to compute the relative event-by-event 
fluctuations of the gluon number~\cite{KNV}.

\medskip

We thank Shigemi Ohta for a helpful discussion.  R. V.'s research was
supported by DOE Contract No. DE-AC02-98CH10886. A.K. and R.V.
acknowledge support from the Portuguese FCT, under grants
CERN/P/FIS/15196/1999 and CERN/P/FIS/40108/2000.
 R.V. and Y.N. thank RIKEN-BNL for support.



\begin{thebibliography}{99}
\bibitem{comment1}The quantity $\Lambda_s$ in this work is identical to 
$g^2\mu$ in Refs.~\cite{AR99,AR00,AR01}. It is approximately twice the 
scale $Q_s$ introduced by A. H. Mueller~\cite{Mueller}.
\bibitem{comment3}Though 
there are potentially gauge artifacts in this method, it was shown 
empirically in Ref.~\cite{AR01} that the results for the integrated number 
per unit rapidity in this method agree well with results from a 
gauge invariant cooling method.
\bibitem{comment4}The number $0.17=f_E\cdot \gamma$. The function $f_E$ is a
slowly varying function of $\Lambda_s R$, varying in the SU(2) case in the 
entire RHIC-LHC range by $20$\%. The function $\gamma$ was shown to be 
constant within error bars for the RHIC-LHC range. Therefore, in the 
regime of interest, we estimate the systematic 
uncertainity of $f_E\cdot \gamma$ to be $~\sim 10$\%.
\bibitem{comment5}We assume that hadronization is isentropic, in accord with 
the ``parton-hadron duality'' seen empirically.
\end{thebibliography}
\end{document}